\documentclass[conference]{IEEEtran}
\IEEEoverridecommandlockouts
\usepackage{cite}
\usepackage{amsmath,amssymb,amsfonts}
\usepackage{algorithmic}
\usepackage{graphicx}
\usepackage{textcomp}
\usepackage{xcolor}
\def\BibTeX{{\rm B\kern-.05em{\sc i\kern-.025em b}\kern-.08em
		T\kern-.1667em\lower.7ex\hbox{E}\kern-.125emX}}

\newtheorem{theorem}{\it Theorem}

\newtheorem{definition}{\it Definition}

\newtheorem{proposition}{\it Proposition}

\title{Relativistic Control:\\Feedback Control of Relativistic Dynamics\\
}

\author{\IEEEauthorblockN{Song Fang and Quanyan Zhu}
	\IEEEauthorblockA{\textit{Department of Electrical and Computer Engineering, New York University, New York, USA}\\
		{\tt \small song.fang@nyu.edu, quanyan.zhu@nyu.edu}
	}
}

\begin{document}

\maketitle

%%%%%%%%%%%%%%%%%%%%%%%%%%%%%%%%%%%%%%%%%%%%%%%%%%%%%%%%%%%%%%%%%%%%%%%%%%%%%%%%
\begin{abstract}
	
Strictly speaking, Newton's second law of motion is only an approximation of the so-called relativistic dynamics, i.e., Einstein's modification of the second law based on his theory of special relativity. Although the approximation is almost exact when the velocity of the dynamical system is far less than the speed of light, the difference will become larger and larger (and will eventually go to infinity) as the velocity approaches the speed of light. Correspondingly, feedback control of such dynamics should also take this modification into consideration (though it will render the system nonlinear), especially when the velocity is relatively large. 
Towards this end, we start this note by studying the state-space representation of the relativistic dynamics. We then investigate on how to employ the feedback linearization approach for such relativistic dynamics, based upon which an additional linear controller may then be designed. As such, the feedback linearization together with the linear controller compose the overall relativistic feedback control law. We also provide discussions on, e.g., controllability, state feedback and output feedback, as well as PID control, in the relativistic setting. 

\end{abstract}

\begin{IEEEkeywords}
	Second law of motion, special relativity, relativistic dynamics, relativistic control, feedback linearization
\end{IEEEkeywords}

%%%%%%%%%%%%%%%%%%%%%%%%%%%%%%%%%%%%%%%%%%%%%%%%%%%%%%%%%%%%%%%%%%%%%%%%%%%%%%%%
\section{Introduction}

Consider a feedback control system where the motion of a dynamical system moving in one dimension is to be controlled. When the velocity $v$ of the moving object is far less than the speed of light $c$ (i.e., when $v \ll c$), which is the case for most such feedback control systems in practical use, Newton's second law of motion \cite{feynman}
$$F = m a $$ 
where $F$ denotes net force, $m$ denotes mass, and $a$ denotes acceleration, holds true almost exactly. On the other hand, Einstein's special relativity \cite{feynman, einstein2013relativity, tolman1917theory, rindler1977essential, rindler2006relativity} indicates why this is only an approximation, which becomes less and less accurate as the moving speed tends to the speed of light. In fact, the relativistic counterpart of Newton's second law should be \cite{rindler2006relativity} 
$$F =  \frac{m_{0} a}{\left(1 - \frac{v^2}{c^2} \right)^{\frac{3}{2}}} $$
where $m_{0}$ denotes the rest mass \cite{rindler2006relativity} (i.e., the mass when $v=0$); this formula truly manifests the underlying relativistic dynamics \cite{feynman} of the system, and it is inherently nonlinear. As such, it can be verified that when $v \ll c$, 
$$F \approx  m_{0} a \left(1 + \frac{3}{2} \frac{v^2}{c^2} \right) \approx  m_{0} a $$
which reduces to the classical Newton's second law. Accordingly, feedback control design for such dynamics should also take this inherent nonlinearity into account, particularly when its velocity is relatively high. 

In this note, we begin by introducing the state-space representation \cite{astrom} of the relativistic dynamics in the one-dimensional case. We then investigate on how to utilize the (exact) feedback linearization \cite{isidori2013nonlinear, khalil2002nonlinear} method for such relativistic dynamics, on top of which an additional linear controller may then be designed. As such, the feedback linearization together with the linear controller compose the overall relativistic feedback control law. We also provide discussions on, e.g., controllability, state feedback and output feedback, as well as PID control, in the relativistic setting. Finally, we examine the three-dimensional case, which is fundamentally different and sophisticated due to the fact that the direction of the acceleration is in general different from that of the net force for three-dimensional relativistic dynamics.

The rest of the note is organized as follows. Section~II introduces Newton's second law of motion and Einstein's modification, i.e., the relativistic dynamics based on the special relativity. In Section~III, we examine the state-space representation of the one-dimensional relativistic dynamics. We also propose the relativistic control design for such relativistic dynamics based on the feedback linearization method. We present discussions on various further topics such as controllability as well. Section~IV is devoted to the three-dimensional case. Concluding remarks are provided in Section~V.

Note that parallel results have been presented in \cite{fang2020relativistic} on the feedback control of relativistic dynamics propelled by ejecting mass rather than an external force, modeling relativistic rocket control or relativistic (space-travel) flight control. 

\section{Preliminaries}

\subsection{Newton's Second Law of Motion and Its State-Space Representation}

For simplicity, we first consider a one-dimensional acceleration system with mass $m \in \mathbb{R} $. Denote its position as $p \left( t \right) \in \mathbb{R}$, its velocity as $ v \left( t \right) \in \mathbb{R}$, and its acceleration as $a \left( t \right) \in \mathbb{R}$. Denote the net force that is acted on the system as $F \left( t \right) \in \mathbb{R}$. 
%For simplicity, we consider a one-dimensional acceleration system without any additional forces such as friction.
According to Newton's second law of motion \cite{feynman}, it holds that
\begin{flalign} \label{Newton}
\ddot{p} \left( t \right) = \dot{v} \left( t \right) = a \left( t \right) = \frac{F \left( t \right)}{m}.
\end{flalign}
In addition, by letting 
\begin{flalign}
\mathbf{x} \left( t \right) = 
\begin{bmatrix} 
p \left( t \right) \\
v \left( t \right) 
\end{bmatrix},~
u \left( t \right) = F \left( t \right),
~
y \left( t \right) = p \left( t \right),
\end{flalign}
the following state-space representation \cite{astrom} of the system may be obtained:
\begin{flalign} \label{newton}
\left\{ \begin{array}{rcl}
\dot{\mathbf{x}} \left( t \right) &=& A \mathbf{x} \left( t \right) + B u \left( t \right), \\
y \left( t \right) &=& C \mathbf{x} \left( t \right),
\end{array}  
\right.
\end{flalign}
where
\begin{flalign} \label{newton2}
A = 
\begin{bmatrix} 
0 & 1 \\
0 & 0 
\end{bmatrix},
~B=\begin{bmatrix} 
0 \\
\frac{1}{m} 
\end{bmatrix},
~C= \begin{bmatrix} 
1 & 0 
\end{bmatrix}.
\end{flalign}

Based on \eqref{newton} and \eqref{newton2}, properties such as controllability as well as different feedback control frameworks including linear state feedback control may then be analyzed or designed \cite{astrom}. 
%In fact, more generally, the linear state feedback control law with reference is given by  $u \left( t \right) = - K_{\mathbf{r}} \mathbf{r} - K \mathbf{x} \left( t \right)$ \cite{astrom}.

%On the other hand, the Luenberger observer for the system
%\begin{flalign}
%\left\{ \begin{array}{rcl}
%\dot{\mathbf{x}} \left( t \right) &=& A \mathbf{x} \left( t \right) + B u \left( t \right), \\
%y \left( t \right) &=& C \mathbf{x} \left( t \right),
%\end{array}  
%\right.
%\end{flalign}
%is given by
%\begin{flalign}
%\dot{\overline{\mathbf{x}}} \left( t \right) 
%= A \overline{\mathbf{x}} \left( t \right) + B u \left( t \right) + L \left( t \right) \left[ y \left( t \right) - C \overline{\mathbf{x}} \left( t \right) \right].
%\end{flalign}
%Define $\mathbf{e} \left( t \right) =  \mathbf{x} \left( t \right) - \overline{\mathbf{x}} \left( t \right)$. Then,
%\begin{flalign}
%\dot{\mathbf{e}} \left( t \right) 
%= \left( A  - LC \right) \mathbf{e} \left( t \right).
%\end{flalign}
%As such, as long as $ A  - LC$ is stable, $\lim\limits_{t \to \infty} \mathbf{e} \left( t \right) = 0$, which means that
%\begin{flalign}
%\lim\limits_{t \to \infty} \overline{\mathbf{x}} \left( t \right) = \lim\limits_{t \to \infty} \mathbf{x} \left( t \right).
%\end{flalign}

\subsection{Relativistic Dynamics}

According to the special relativity \cite{rindler2006relativity}, however, \eqref{Newton} is only an approximation. Its relativistic modification will be introduced as follows.

Still consider a one-dimensional acceleration system. Denote its position as $p \left( t \right) \in \mathbb{R}$, its velocity as $ v \left( t \right) \in \mathbb{R}$, and its acceleration as $a \left( t \right) \in \mathbb{R}$. Denote the net force that is acted on the system as $F \left( t \right) \in \mathbb{R}$. 
According to the special relativity, Newton's second law of motion should be modified as \cite{rindler2006relativity}
\begin{flalign} \label{RNewton}
\ddot{p} \left( t \right) 
= \dot{v} \left( t \right) 
= a \left( t \right) 
= \frac{F \left( t \right)}{m_{0}} \left[ 1 - \frac{v^2 \left( t \right)}{c^2} \right]^{\frac{3}{2}}.
\end{flalign}
Herein, $c$ denotes the speed of light, while $m_{0}$ denotes the rest mass \cite{rindler2006relativity}, i.e., the mass when $v \left( t \right) = 0$.

A detailed derivation of \eqref{RNewton} can be found in, e.g., \cite{rindler2006relativity}. It is clear that \eqref{RNewton} is different from \eqref{Newton} by a factor of 
\begin{flalign}
\left[ 1 - \frac{v^2 \left( t \right)}{c^2} \right]^{\frac{3}{2}},
\end{flalign} 
which in fact renders the system of \eqref{RNewton} nonlinear; note that the system is \eqref{Newton} is linear. 
Moreover, when $v \left( t \right) \ll c$, 
\begin{flalign}
F \left( t \right) 
&= m_{0} a \left( t \right) \left[ 1 - \frac{v^2 \left( t \right)}{c^2} \right]^{-\frac{3}{2}}
\approx  m_{0} a \left( t \right) \left[ 1 + \frac{3}{2} \frac{v^2 \left( t \right)}{c^2} \right] \nonumber \\
&\approx  m_{0} a \left( t \right),
\end{flalign}
which reduces to \eqref{Newton} \cite{rindler2006relativity}.
On the other hand, when $v \left( t \right) \to c$,
\begin{flalign}
\left[ 1 - \frac{v^2 \left( t \right)}{c^2} \right]^{ - \frac{3}{2}} \to \infty,
\end{flalign}  
meaning that it will take infinite force to accelerate when the velocity approaches the speed of light \cite{rindler2006relativity}, i.e., the difference between \eqref{RNewton} and \eqref{Newton} will tend to infinity in this extreme case.

\section{Feedback Control of Relativistic Dynamics}

In this section, we discuss the state-space representation and feedback control of relativistic dynamics. 
We first introduce the state-space representation of \eqref{RNewton}.

\subsection{State-Space Representation of Relativistic Dynamics}

By letting
\begin{flalign}
\mathbf{x} \left( t \right) = 
\begin{bmatrix} 
p \left( t \right) \\
v \left( t \right) 
\end{bmatrix},~u \left( t \right) = F \left( t \right),~y \left( t \right) = p \left( t \right),
\end{flalign}
we obtain the following state-space representation of \eqref{RNewton}:
\begin{flalign} \label{ss1}
\left\{ \begin{array}{rcl}
\dot{\mathbf{x}} \left( t \right) &=& f \left[  \mathbf{x} \left( t \right), u \left( t \right) \right], \\
y \left( t \right) &=& C \mathbf{x} \left( t \right),
\end{array}  
\right.
\end{flalign}
where
\begin{flalign}
f \left[  \mathbf{x} \left( t \right), u \left( t \right) \right]
=\begin{bmatrix} 
\begin{bmatrix} 
0 & 1
\end{bmatrix} \mathbf{x} \left( t \right) \\
\frac{u \left( t \right)}{m_{0}} \left\{ 1- \frac{\left[ \begin{bmatrix} 
	0 & 1
	\end{bmatrix} \mathbf{x} \left( t \right) \right]^2}{c^2} \right\}^{\frac{3}{2}}  
\end{bmatrix},
\end{flalign}
and
\begin{flalign}
C= \begin{bmatrix} 
1 & 0 
\end{bmatrix}.
\end{flalign}
This can be equivalently rewritten as
\begin{flalign} \label{Rstate1}
\left\{ \begin{array}{rcl}
\dot{\mathbf{x}} \left( t \right) &=& A \mathbf{x} \left( t \right) + B \left\{ 1- \frac{\left[ \begin{bmatrix} 
	0 & 1
	\end{bmatrix} \mathbf{x} \left( t \right) \right]^2}{c^2} \right\}^{\frac{3}{2}} u \left( t \right), \\
y \left( t \right) &=& C \mathbf{x} \left( t \right),
\end{array}  
\right.
\end{flalign}
where
\begin{flalign} \label{Rstate2}
A = 
\begin{bmatrix} 
0 & 1 \\
0 & 0 
\end{bmatrix},
~B=\begin{bmatrix} 
0 \\
\frac{1}{m_{0}} 
\end{bmatrix},
~C= \begin{bmatrix} 
1 & 0 
\end{bmatrix},
\end{flalign}
and it may be viewed as the state-space representation of the relativistic dynamics.

We summarize the previous discussions of this subsection in the following proposition.

\begin{proposition}
    The state-space representation of the relativistic dynamics \eqref{RNewton} is given by \eqref{Rstate1} and \eqref{Rstate2}.
\end{proposition}

%As a matter of fact, we have negel

%If we include the mass $m \left( t \right)$ in the state as well, that is, if we let
%\begin{flalign}
%\mathbf{x} \left( t \right) = 
%\begin{bmatrix} 
%p \left( t \right) \\
%v \left( t \right) \\
%m \left( t \right) 
%\end{bmatrix},~u \left( t \right) = F \left( t \right),~y \left( t \right) = p \left( t \right),
%\end{flalign} 
%then in \eqref{ss1}, we have
%\begin{flalign}
%f \left[  \mathbf{x} \left( t \right), u \left( t \right) \right]
%=\begin{bmatrix} 
%\begin{bmatrix} 
%0 & 1 & 0
%\end{bmatrix} \mathbf{x} \left( t \right) \\
%\frac{u \left( t \right)}{\begin{bmatrix} 
%		0 & 0 & 1 \end{bmatrix} \mathbf{x} \left( t \right)} \left\{ 1- \frac{\left[ \begin{bmatrix} 
%	0 & 1 & 0
%	\end{bmatrix} \mathbf{x} \left( t \right) \right]^2}{c^2} \right\}  \\
%\left\{ 1- \frac{\left[ \begin{bmatrix} 
%	0 & 1 & 0
%	\end{bmatrix} \mathbf{x} \left( t \right) \right]^2}{c^2} \right\}^{\frac{3}{2}}   
%\end{bmatrix},
%\end{flalign}
%and
%\begin{flalign}
%C= \begin{bmatrix} 
%1 & 0 
%\end{bmatrix}.
%\end{flalign}

\subsection{Feedback Linearization of Relativistic Dynamics}

We will now linearize \cite{isidori2013nonlinear, khalil2002nonlinear} the system of \eqref{Rstate1} and \eqref{Rstate2} with the following transformation:
\begin{flalign} \label{linearize1}
w \left( t \right) 
=  \left\{ 1- \frac{\left[ \begin{bmatrix} 
	0 & 1
	\end{bmatrix} \mathbf{x} \left( t \right) \right]^2}{c^2} \right\}^{\frac{3}{2}} u \left( t \right).
\end{flalign}
It can be verified that the transformation in \eqref{linearize1} is invertible for $\left| v \left( t \right) \right| = \left| \begin{bmatrix} 
0 & 1
\end{bmatrix} \mathbf{x} \left( t \right) \right| < c$, and its inverse is given by
\begin{flalign} \label{linearize2}
u \left( t \right)
=  \left\{ 1- \frac{\left[ \begin{bmatrix} 
	0 & 1
	\end{bmatrix} \mathbf{x} \left( t \right) \right]^2}{c^2} \right\}^{- \frac{3}{2}} w \left( t \right).
\end{flalign}
In addition, both \eqref{linearize1} and \eqref{linearize2} are continuously differentiable for $\left| v \left( t \right) \right|  < c$. As such, the transformation in \eqref{linearize1} is a diffeomorphism and the system of \eqref{Rstate1} and \eqref{Rstate2} is thus feedback linearizable \cite{isidori2013nonlinear, khalil2002nonlinear} for $\left| v \left( t \right) \right| < c$.

On the other hand, it is known from the special relativity \cite{rindler2006relativity} that $\left| v \left( t \right) \right| < c$ will always hold, i.e., this is inherent in the relativistic dynamics. In this sense, the system is always relativistically feedback linearizable, and we may always linearize the system using \eqref{linearize1} and \eqref{linearize2}.

As a matter of fact, with transformations \eqref{linearize1} and \eqref{linearize2}, the system from $w \left( t \right)$ to $y \left( t \right)$ will then be linear with its state-space model given by
\begin{flalign} \label{linearized1}
\left\{ \begin{array}{rcl}
\dot{\mathbf{x}} \left( t \right) &=& A \mathbf{x} \left( t \right) + B w \left( t \right), \\
y \left( t \right) &=& C \mathbf{x} \left( t \right),
\end{array}  
\right.
\end{flalign}
where
\begin{flalign} \label{linearized2}
A = 
\begin{bmatrix} 
0 & 1 \\
0 & 0 
\end{bmatrix},
~B=\begin{bmatrix} 
0 \\
\frac{1}{m_{0}} 
\end{bmatrix},
~C= \begin{bmatrix} 
1 & 0 
\end{bmatrix}.
\end{flalign}

We now summarize the previous discussions of this subsection in the following theorem.

\begin{theorem}
	If we choose $w \left( t \right)$ as in \eqref{linearize1} and \eqref{linearize2}, then the system of \eqref{Rstate1} and \eqref{Rstate2} can be linearized as \eqref{linearized1} and \eqref{linearized2}.
\end{theorem}

As such, we may first design a controller for the linearized dynamics \eqref{linearized1} and \eqref{linearized2}. Then, this controller together with the feedback linearization compose the overall feedback controller design for the relativistic dynamics of \eqref{Rstate1} and \eqref{Rstate2}. Detailed discussions on this as well as properties such as controllability will be presented in what follows.

\subsection{Relativistic Control}

In this subsection, we provide a series of discussions concerning relativistic control.

%
%\subsubsection{Regulation or Tracking (Velocity and/or Position)}
%
%\subsubsection{Explicit Expressions}

\subsubsection{Relativistic Controllability}

Although the system of \eqref{newton} and \eqref{newton2} is controllable, the system of \eqref{Rstate1} and \eqref{Rstate2} is not controllable \cite{astrom, dorf2011modern}, since any state $\mathbf{x} \left( t \right)$ with $\left| v \left( t \right) \right| > c$ is not reachable. 
Strictly speaking, any motion system with rest mass $m_0 > 0$ is not controllable, if velocity is included in the state, due to the fact that its velocity cannot exceed the speed of light according to the special relativity \cite{rindler2006relativity}.

We next propose the notion of relativistic controllability to differentiate whether the system is controllable or not for $\left| v \left( t \right) \right| < c$.

\begin{definition}
	We say the system is relativistically controllable if the system input $u \left( t \right): \left[ t_0,~T \right] \to \mathbb{R}$ can be designed to steer any initial state $\mathbf{x} \left( t_0 \right)$ with $\left| v  \left( t \right) \right| < c$ to any other final state $\mathbf{x} \left( T \right)$ with $\left| v  \left( t \right) \right| < c$ in a finite time interval $T < \infty$.
\end{definition}

It may then be verified that although the system given in \eqref{Rstate1} and \eqref{Rstate2} is not controllable, it is relativistically controllable. To see this, note first that the system given in \eqref{linearized1} and \eqref{linearized2} is controllable. Hence, the system input $w \left( t \right): \left[ t_0,~T \right] \to \mathbb{R}$ can be designed to steer any initial state $\mathbf{x} \left( t_0 \right)$ to any other final state $\mathbf{x} \left( T \right)$ in a finite time. As a special case, it will certainly be possible to design a $w \left( t \right): \left[ t_0,~T \right] \to \mathbb{R}$ to steer any initial state $\mathbf{x} \left( t_0 \right)$ with $\left| v  \left( t \right) \right| < c$ to any other final state $\mathbf{x} \left( T \right)$ with $\left| v  \left( t \right) \right| < c$ in a finite time. Correspondingly, with $u \left( t \right): \left[ t_0,~T \right] \to \mathbb{R}$, where
\begin{flalign} 
u \left( t \right)
=  \left[ 1- \frac{\left| v \left( t \right) \right|^2}{c^2} \right]^{- \frac{3}{2}} w \left( t \right),~\left| v  \left( t \right) \right| < c,
\end{flalign}
any initial state $\mathbf{x} \left( t_0 \right)$ with $\left| v  \left( t \right) \right| < c$ can thus be steered to any other final state $\mathbf{x} \left( T \right)$ with $\left| v  \left( t \right) \right| < c$ in a finite time for the system given by \eqref{Rstate1} and \eqref{Rstate2}.

On the other hand, relativistic observability may be defined and analyzed in a similar spirit.

%On the other hand, we also propose the notion of relativistic observability to differentiate whether the system is observable for $\left| v \left( t \right) \right| < c$.
%
%\begin{definition}
%	We say the system is relativistically observable if all the states $\mathbf{x} \left( t \right)$ with $\left| v \left( t \right) \right| < c$ can be observed.
%\end{definition}

\subsubsection{Relativistic State Feedback}

If the state feedback controller for the system of \eqref{linearized1} and \eqref{linearized2} is designed as \cite{astrom}
\begin{flalign}
w \left( t \right)
= - K  \mathbf{x} \left( t \right),
\end{flalign}
then, noting also that
\begin{flalign}
u \left( t \right)
=  \left\{ 1- \frac{\left[ \begin{bmatrix} 
	0 & 1
	\end{bmatrix} \mathbf{x} \left( t \right) \right]^2}{c^2} \right\}^{- \frac{3}{2}} w \left( t \right),
\end{flalign}
the overall relativistic state feedback controller is given by
\begin{flalign} \label{Rstatefeedback}
u \left( t \right)
=  - \left\{ 1- \frac{\left[ \begin{bmatrix} 
	0 & 1
	\end{bmatrix} \mathbf{x} \left( t \right) \right]^2}{c^2} \right\}^{- \frac{3}{2}} K  \mathbf{x} \left( t \right).
\end{flalign}

We summarize the discussions on state feedback in the following definition.

\begin{definition}
	The relativistic state feedback controller is given by \eqref{Rstatefeedback}.
\end{definition}

%More generally, reference?

\subsubsection{Relativistic Output Feedback}

If the output feedback controller for the system of \eqref{linearized1} and \eqref{linearized2} is designed as \cite{astrom}
\begin{flalign}
w \left( t \right)
= l \left[ y \left( t \right) \right],
\end{flalign}
then, noting as well that
\begin{flalign}
u \left( t \right)
= \left\{ 1- \frac{\left[ \begin{bmatrix} 
	0 & 1
	\end{bmatrix} \mathbf{x} \left( t \right) \right]^2}{c^2} \right\}^{- \frac{3}{2}} w \left( t \right),
\end{flalign}
with
\begin{flalign}
\begin{bmatrix} 
0 & 1
\end{bmatrix} \mathbf{x} \left( t \right) 
= \dot{ y } \left( t \right),
\end{flalign}  
the overall relativistic output-feedback controller is given by
\begin{flalign} \label{output}
u \left( t \right)
=  \left[ 1- \frac{\dot{ y }^2 \left( t \right)}{c^2} \right]^{- \frac{3}{2}} l \left[ y \left( t \right) \right].
\end{flalign}

We next summarize the discussions concerning output feedback in the following definition.

\begin{definition}
	The relativistic output feedback controller is given by \eqref{output}.
\end{definition}

\subsubsection{Relativistic PID Control}

We now consider a special case of output feedback: PID control. Suppose that $w \left( t \right)$ is designed using PID control as \cite{astrom}
\begin{flalign}
w(t)=K_{\text{p}} e \left( t \right) + K_{\text{i}} \int_{0}^{t}e \left( \tau \right) \text{d}\tau+K_{\text{d}} \dot{ e } \left( t \right),
\end{flalign}
where 
\begin{flalign} 
e \left( t \right) = r - y \left( t \right).
\end{flalign} 
Note that herein $r$ is given in the controller frame; see discussions in Section~\ref{frame} for more details.
Meanwhile, 
\begin{flalign}
\begin{bmatrix} 
0 & 1
\end{bmatrix} \mathbf{x} \left( t \right) = \dot{ y } \left( t \right) =  - \dot{ e } \left( t \right).
\end{flalign}  
As such, noting also that
\begin{flalign}
u \left( t \right)
= \left\{ 1- \frac{\left[ \begin{bmatrix} 
	0 & 1
	\end{bmatrix} \mathbf{x} \left( t \right) \right]^2}{c^2} \right\}^{- \frac{3}{2}} w \left( t \right),
\end{flalign}
the overall relativistic PID controller is given by
\begin{flalign} \label{RPID}
u \left( t \right) =  \left[ 1- \frac{\dot{ e }^2 \left( t \right)}{c^2} \right]^{- \frac{3}{2}} \left[ K_{\text{p}} e \left( t \right) + K_{\text{i}} \int_{0}^{t}e \left( \tau \right) \text{d}\tau+K_{\text{d}} \dot{ e } \left( t \right) \right].
\end{flalign}

We summarize the discussions about PID control in the following definition.

\begin{definition}
	The relativistic PID controller is given by \eqref{RPID}.
\end{definition}

Note that herein we assumed that $r$ is a constant and thus $ \dot{ r } = 0$. If this is not the case, we can invoke the fact
\begin{flalign}
\begin{bmatrix} 
0 & 1
\end{bmatrix} \mathbf{x} \left( t \right) = \dot{ y } \left( t \right) = \dot{ r } \left( t \right)  - \dot{ e } \left( t \right),
\end{flalign} 
and obtain a similar result.

%Strictly speaking, PID control belongs to output feedback control, which we will discuss in Section~\ref{}.

\subsubsection{Controller Frame vs. Plant Frame} \label{frame}

It is worth emphasizing that in this note, the time $t$ as well as $a \left( t \right) $, $v \left( t \right)$, $p \left( t \right) $, $F \left( t \right) $ and so on are all defined and measured in the controller frame, and the analysis and design are carried out in the controller frame as well \cite{rindler2006relativity}. As such, the reference signal $r$ shall also be given in the controller frame. On the other hand, in the plant frame, the reference for position $r_{p \left( t \right)}$ will be transformed into 
\begin{flalign}
\left[ 1- \frac{v^2 \left( t \right) }{c^2} \right]^{- \frac{1}{2}} r_{p \left( t \right)},
\end{flalign}
due to length contraction \cite{rindler2006relativity}, while the reference for velocity $r_{v \left( t \right)}$ stays the same \cite{rindler2006relativity}. We will, however, leave detailed discussions on this topic to future research.

\section{The Three-Dimensional Case}
 
For simplicity and clarity of the presentation, we separate the three-dimensional case completely from the previous one-dimensional case, and we collect all the discussions concerning the three-dimensional case into this individual section. It will be seen that although the results for the three-dimensional case are organized in a way similar to that of the one-dimensional case, they are not trivial extensions and the implications (and accordingly the feedback control design as well) are more sophisticated. In particular, the direction of the acceleration is in general not the same as that of the net force in three-dimensional relativistic dynamics, which is fundamentally different from the classical Newton's second law of motion in the three-dimensional case; for more details on this as well as how it might affect the overall control design, refer to the theoretical deviations in what follows.

\subsection{Newton's Second Law of Motion and Its State-Space Representation}

We now consider a three-dimensional acceleration system with mass $m \in \mathbb{R}$. Denote its position as $\mathbf{p} \left( t \right) \in \mathbb{R}^3$, its velocity as $\mathbf{v} \left( t \right) \in \mathbb{R}^3$, and its acceleration as $\mathbf{a} \left( t \right) \in \mathbb{R}^3$. Denote the net force that is acted on the system as $\mathbf{F} \left( t \right) \in \mathbb{R}^3$.
%For simplicity, we consider a one-dimensional acceleration system without any additional forces such as friction.
According to Newton's second law of motion \cite{feynman}, it holds that
\begin{flalign} \label{Newton3}
\ddot{\mathbf{p}} \left( t \right) = \dot{\mathbf{v}} \left( t \right) = \mathbf{a} \left( t \right) = \frac{\mathbf{F}  \left( t \right)}{m}.
\end{flalign}
It is clear that direction of $\mathbf{a} \left( t \right)$ will always be the same as that of $\mathbf{F}  \left( t \right)$.
In addition, by letting
\begin{flalign}
\mathbf{x} \left( t \right) = 
\begin{bmatrix} 
\mathbf{p} \left( t \right) \\
\mathbf{v} \left( t \right) 
\end{bmatrix},
~
\mathbf{u} \left( t \right) = \mathbf{F} \left( t \right),
~
\mathbf{y} \left( t \right) = \mathbf{p} \left( t \right),
\end{flalign}
the state-space representation of the system may be obtained as
\begin{flalign} \label{newton3}
\left\{ \begin{array}{rcl}
\dot{\mathbf{x}} \left( t \right) &=& A \mathbf{x} \left( t \right) + B \mathbf{u} \left( t \right), \\
\mathbf{y} \left( t \right) &=& C \mathbf{x} \left( t \right),
\end{array}  
\right.
\end{flalign}
where
\begin{flalign} \label{newton4}
A = 
\begin{bmatrix} 
0_3 & I_3 \\
0_3 & 0_3 
\end{bmatrix},
~B=\begin{bmatrix} 
0_3 \\
\frac{1}{m} I_3 
\end{bmatrix},
~C= \begin{bmatrix} 
I_3 & 0_3 
\end{bmatrix}.
\end{flalign}
Herein, $0_3 $ denotes the zero matrix of dimension $3 \times 3$ while $I_3 $ denotes the identity matrix of dimension $3 \times 3$.

\subsection{Relativistic Dynamics}

Again, \eqref{Newton3} is only an approximation. Its relativistic modification will be presented as follows.

Still consider a three-dimensional acceleration system. Denote its position as $\mathbf{p} \left( t \right) \in \mathbb{R}^3$, its velocity as $\mathbf{v} \left( t \right) \in \mathbb{R}^3$, and its acceleration as $\mathbf{a} \left( t \right) \in \mathbb{R}^3$. Denote the net force that is acted on the system as $\mathbf{F} \left( t \right) \in \mathbb{R}^3$.
In addition, let $c$ denote the speed of light, and let $m_{0}$ denote the rest mass, i.e., the mass when $v \left( t \right) = 0$. According to the special relativity, Newton's second law of motion should be modified as \cite{rindler2006relativity}
\begin{flalign} \label{RNewton3}
\ddot{\mathbf{p}} \left( t \right) 
&= \dot{\mathbf{v}} \left( t \right) 
= \mathbf{a} \left( t \right) \nonumber \\
&= {\frac{\left[ 1 - \frac{\left| \mathbf{v} \left( t \right) \right|^2}{c^2} \right]^{\frac{1}{2}}}{m_{0}}}\left\{\mathbf{F} \left( t \right) - {\frac{ \left[ \mathbf{v} \left( t \right) \cdot \mathbf{F} \left( t \right) \right] \mathbf{v} \left( t \right) }{c^{2}}}\right\},
\end{flalign}
or equivalently,
\begin{flalign} \label{RNewton32}
\mathbf{F} \left( t \right) 
&= \frac{m_{0}}{\left[ 1 - \frac{\left| \mathbf{v} \left( t \right) \right|^2}{c^2} \right]^{\frac{3}{2}}}{\frac{\left[ \mathbf{v} \left( t \right) \cdot \mathbf{a} \left( t \right) \right] \mathbf{v} \left( t \right) }{c^{2}}} \nonumber \\
&~~~~ + \frac{m_{0}}{\left[ 1 - \frac{\left| \mathbf{v} \left( t \right) \right|^2}{c^2} \right]^{\frac{1}{2}}} \mathbf{a} \left( t \right).
\end{flalign}

It is then implicated \cite{rindler2006relativity} that while $ \mathbf{a} \left( t \right) $ is always coplanar with $\mathbf{F} \left( t \right)$ and $\mathbf{v} \left( t \right)$, it is in general not in the same direction as that of $\mathbf{F} \left( t \right)$. In fact, if $\mathbf{F} \left( t \right)$ is splitted into a component $\mathbf{F}_{\parallel} \left( t \right)$ parallel to $\mathbf{v} \left( t \right)$ and the other $\mathbf{F}_{\perp} \left( t \right)$ orthogonal to $\mathbf{v} \left( t \right)$ while $\mathbf{a} \left( t \right)$ is also splitted into a component $\mathbf{a}_{\parallel} \left( t \right)$ parallel to $\mathbf{v} \left( t \right)$ and the other $\mathbf{a}_{\perp} \left( t \right)$ orthogonal to $\mathbf{v} \left( t \right)$, then it can be verified \cite{rindler2006relativity} that
\begin{flalign} 
\mathbf{F}_{\parallel} \left( t \right) 
= \frac{m_{0}}{\left[ 1 - \frac{\left| \mathbf{v} \left( t \right) \right|^2}{c^2} \right]^{\frac{3}{2}}} \mathbf{a}_{\parallel} \left( t \right),
\end{flalign}
while
\begin{flalign} 
\mathbf{F}_{\perp} \left( t \right) 
= \frac{m_{0}}{\left[ 1 - \frac{\left| \mathbf{v} \left( t \right) \right|^2}{c^2} \right]^{\frac{1}{2}}} \mathbf{a}_{\perp} \left( t \right).
\end{flalign}
This means that it is as if the system manifests different inertial resistances (different ``masses"; see discussions in \cite{rindler2006relativity}) to the same force in different directions, depending on whether it is subjected to that force longitudinally or transversely.

On the other hand, it can be verified that when $\left| \mathbf{v} \left( t \right) \right| \ll c$, \eqref{RNewton32} reduces to \eqref{Newton3} \cite{rindler2006relativity}.

\subsection{State-Space Representation of Relativistic Dynamics}

As such, if we let
\begin{flalign}
\mathbf{x} \left( t \right) = 
\begin{bmatrix} 
\mathbf{p} \left( t \right) \\
\mathbf{v} \left( t \right) 
\end{bmatrix},
~
\mathbf{u} \left( t \right) = \mathbf{F} \left( t \right),
~
\mathbf{y} \left( t \right) = \mathbf{p} \left( t \right),
\end{flalign}
we may rewrite \eqref{RNewton3} and \eqref{RNewton32} as
\begin{flalign}
\left\{ \begin{array}{rcl}
\dot{\mathbf{x}} \left( t \right) &=& \mathbf{f} \left[  \mathbf{x} \left( t \right), \mathbf{u} \left( t \right) \right], \\
\mathbf{y} \left( t \right) &=& C \mathbf{x} \left( t \right),
\end{array}  
\right.
\end{flalign}
where $\mathbf{f} \left[  \mathbf{x} \left( t \right), \mathbf{u} \left( t \right) \right]$ is given by \eqref{long} (for the long equations such as \eqref{long}, see the last page of this note)
\begin{figure*}
	%\begin{strip}
	%\noindent\rule{\textwidth}{0.4pt}
	\begin{flalign} \label{long}
	\mathbf{f} \left[  \mathbf{x} \left( t \right), \mathbf{u} \left( t \right) \right]
	=\begin{bmatrix} 
	\begin{bmatrix} 
	0_3 & I_3
	\end{bmatrix} \mathbf{x} \left( t \right) \\
	{\frac{ \left[ 1 - \frac{\left| \begin{bmatrix} 
				0_3 & I_3
				\end{bmatrix} \mathbf{x} \left( t \right) \right|^2 }{c^2} \right]^{\frac{1}{2}}}{m_{0}}}\left\{\mathbf{u} \left( t \right) - {\frac{ \left\{ \left[ \begin{bmatrix} 
			0_3 & I_3
			\end{bmatrix} \mathbf{x} \left( t \right) \right] \cdot \mathbf{u} \left( t \right) \right\} \begin{bmatrix} 
			0_3 & I_3
			\end{bmatrix} \mathbf{x} \left( t \right) }{c^{2}}}\right\}  
	\end{bmatrix}
	\end{flalign}
	%\noindent\rule{\textwidth}{0.4pt}
	%\end{strip}
\end{figure*}
and
\begin{flalign}
C= \begin{bmatrix} 
I_3 & 0_3 
\end{bmatrix}.
\end{flalign}
This is equivalent to
\begin{figure*}
	%\begin{strip}
	%\noindent\rule{\textwidth}{0.4pt}
	\begin{flalign} \label{long4}
	\left\{ \begin{array}{rcl}
	\dot{\mathbf{x}} \left( t \right) &=& A \mathbf{x} \left( t \right) + B { \left[ 1 - \frac{\left| \begin{bmatrix} 
			0_3 & I_3
			\end{bmatrix} \mathbf{x} \left( t \right) \right|^2 }{c^2} \right]^{\frac{1}{2}}}\left\{\mathbf{u} \left( t \right) - {\frac{ \left\{ \left[ \begin{bmatrix} 
			0_3 & I_3
			\end{bmatrix} \mathbf{x} \left( t \right) \right] \cdot \mathbf{u} \left( t \right) \right\} \begin{bmatrix} 
			0_3 & I_3
			\end{bmatrix} \mathbf{x} \left( t \right) }{c^{2}}}\right\} \\
	\mathbf{y} \left( t \right) &=& C \mathbf{x} \left( t \right)
	\end{array}  
	\right.
	\end{flalign}
	%\noindent\rule{\textwidth}{0.4pt}
	%\end{strip}
\end{figure*}
\eqref{long4} (see the last page),
where
\begin{flalign} \label{long4m}
A = 
\begin{bmatrix} 
0_3 & I_3 \\
0_3 & 0_3 
\end{bmatrix},
~B=\begin{bmatrix} 
0_3 \\
\frac{1}{m_{0}} I_3
\end{bmatrix},
~C= \begin{bmatrix} 
I_3 & 0_3 
\end{bmatrix},
\end{flalign}
and it may be viewed as the state-space representation of the relativistic dynamics.

We next summarize the discussions of this subsection in the following proposition.

\begin{proposition}
	The state-space representation of the relativistic dynamics \eqref{RNewton3} is given by \eqref{long4} and \eqref{long4m}.
\end{proposition}

\subsection{Feedback Linearization of Relativistic Dynamics}

We will now linearize \cite{isidori2013nonlinear, khalil2002nonlinear} the system of \eqref{long4} and \eqref{long4m} using the transformation given by \eqref{long2} (see the last page).
\begin{figure*}
	\begin{flalign} \label{long2}
	\mathbf{w} \left( t \right) 
	= {\frac{ \left[ 1 - \frac{\left| \begin{bmatrix} 
				0_3 & I_3
				\end{bmatrix} \mathbf{x} \left( t \right) \right|^2 }{c^2} \right]^{\frac{1}{2}}}{m_{0}}}\left\{\mathbf{u} \left( t \right) - {\frac{ \left\{ \left[ \begin{bmatrix} 
			0_3 & I_3
			\end{bmatrix} \mathbf{x} \left( t \right) \right] \cdot \mathbf{u} \left( t \right) \right\} \begin{bmatrix} 
			0_3 & I_3
			\end{bmatrix} \mathbf{x} \left( t \right) }{c^{2}}}\right\}
	\end{flalign}
%	\noindent\rule{\textwidth}{0.4pt}
\end{figure*}
\begin{figure*}
	\begin{flalign} \label{long3}
	\mathbf{u} \left( t \right) 
	= \frac{m_{0}}{\left[ 1 - \frac{\left| \begin{bmatrix} 
			0_3 & I_3
			\end{bmatrix} \mathbf{x} \left( t \right) \right|^2 }{c^2} \right]^{\frac{3}{2}}}{\frac{\left\{ \left[ \begin{bmatrix} 
			0_3 & I_3
			\end{bmatrix} \mathbf{x} \left( t \right) \right] \cdot \mathbf{w} \left( t \right) \right\} \begin{bmatrix} 
			0_3 & I_3
			\end{bmatrix} \mathbf{x} \left( t \right) }{c^{2}}} + \frac{m_{0}}{\left[ 1 - \frac{\left| \begin{bmatrix} 
			0_3 & I_3
			\end{bmatrix} \mathbf{x} \left( t \right) \right|^2 }{c^2} \right]^{\frac{1}{2}}} \mathbf{w} \left( t \right)
	\end{flalign}
	%\noindent\rule{\textwidth}{0.4pt}
\end{figure*}
It can be verified that \eqref{long2} is invertible, and its inverse is given by \eqref{long3} (see the last page).
In addition, both \eqref{long2} and \eqref{long3} are continuously differentiable for $\left| \mathbf{v} \left( t \right) \right| < c$. As such, the transformation in \eqref{long2} is a diffeomorphism and the system of \eqref{long4} and \eqref{long4m} is thus feedback linearizable \cite{isidori2013nonlinear, khalil2002nonlinear} for $\left| \mathbf{v} \left( t \right) \right| < c$.
On the other hand, it is known from the special relativity \cite{rindler2006relativity} that $\left| \mathbf{v} \left( t \right) \right| < c$ will always hold, i.e., this is inherent in the relativistic dynamics. In this sense, the system is always relativistically feedback linearizable, and we may always linearize the system using \eqref{long2} and \eqref{long3}.
In fact, with \eqref{long2} and \eqref{long3}, the system from $\mathbf{w} \left( t \right)$ to $\mathbf{y} \left( t \right)$ will then be linear with its state-space model given by
\begin{flalign} \label{linearized31}
\left\{ \begin{array}{rcl}
\dot{\mathbf{x}} \left( t \right) &=& A \mathbf{x} \left( t \right) + B \mathbf{w} \left( t \right), \\
\mathbf{y} \left( t \right) &=& C \mathbf{x} \left( t \right),
\end{array}  
\right.
\end{flalign}
where
\begin{flalign} \label{linearized32}
A = 
\begin{bmatrix} 
0_3 & I_3 \\
0_3 & 0_3
\end{bmatrix},
~B=\begin{bmatrix} 
0_3 \\
\frac{1}{m_{0}} I_3 
\end{bmatrix},
~C= \begin{bmatrix} 
I_3 & 0_3
\end{bmatrix}.
\end{flalign}

We now summarize the previous discussions of this subsection in the following theorem.

\begin{theorem}
	If we choose $\mathbf{w} \left( t \right)$ as in \eqref{long2} and \eqref{long3}, then the system of \eqref{long4} and \eqref{long4m} can be linearized as \eqref{linearized31} and \eqref{linearized32}.
\end{theorem}

It is worth mentioning that if $\mathbf{u} \left( t \right)$ is splitted into a component $\mathbf{u}_{\parallel} \left( t \right)$ parallel to $\begin{bmatrix} 
0_3 & I_3
\end{bmatrix} \mathbf{x} \left( t \right)$ and the other $\mathbf{u}_{\perp} \left( t \right)$ orthogonal to $\begin{bmatrix} 
0_3 & I_3
\end{bmatrix} \mathbf{x} \left( t \right)$ while $\mathbf{w} \left( t \right)$ is also splitted into a component $\mathbf{w}_{\parallel} \left( t \right)$ parallel to $\begin{bmatrix} 
0_3 & I_3
\end{bmatrix} \mathbf{x} \left( t \right)$ and the other $\mathbf{w}_{\perp} \left( t \right)$ orthogonal to $\begin{bmatrix} 
0_3 & I_3
\end{bmatrix} \mathbf{x} \left( t \right)$, then it holds that
\begin{flalign} 
\mathbf{u}_{\parallel} \left( t \right) 
= \frac{m_{0}}{\left[ 1 - \frac{\left| \begin{bmatrix} 
		0_3 & I_3
		\end{bmatrix} \mathbf{x} \left( t \right) \right|^2}{c^2} \right]^{\frac{3}{2}}} \mathbf{w}_{\parallel} \left( t \right),
\end{flalign}
while
\begin{flalign} 
\mathbf{u}_{\perp} \left( t \right) 
= \frac{m_{0}}{\left[ 1 - \frac{\left| \begin{bmatrix} 
		0_3 & I_3
		\end{bmatrix} \mathbf{x} \left( t \right) \right|^2}{c^2} \right]^{\frac{1}{2}}} \mathbf{w}_{\perp} \left( t \right).
\end{flalign}
As such, it is as if the relativistic feedback linearization is composed of different transformations longitudinally and transversely, as a result of the fact that the ``longitudinal mass" and ``transverse mass" of the system are different \cite{rindler2006relativity}.

\subsection{Relativistic Control}

We may then analyze the properties of the system of \eqref{long4} and \eqref{long4m} and design the feedback controller for it as in the one-dimensional case. For instance, it can be verified the system of \eqref{long4} and \eqref{long4m} is not controllable (though the system of \eqref{newton3} and \eqref{newton4} is controllable), but it is relativistically controllable, whereas relativistic controllability in the general case is defined as follows.

\begin{definition}
	We say the system is relativistically controllable if the system input $\mathbf{u} \left( t \right): \left[ t_0,~T \right] \to \mathbb{R}^{n}$ can be designed to steer any initial state $\mathbf{x} \left( t_0 \right)$ with $\left| \mathbf{v}  \left( t \right) \right| < c$ to any other final state $\mathbf{x} \left( T \right)$ with $\left| \mathbf{v} \left( t \right) \right| < c$ in a finite time.
\end{definition}

For another two examples, the relativistic state feedback controller may be obtained as \eqref{long5} (see the last page),
\begin{figure*}
	\begin{flalign} \label{long5}
	\mathbf{u} \left( t \right) 
	= \frac{m_{0}}{\left[ 1 - \frac{\left| \begin{bmatrix} 
			0_3 & I_3
			\end{bmatrix} \mathbf{x} \left( t \right) \right|^2 }{c^2} \right]^{\frac{3}{2}}}{\frac{\left\{ \left[ \begin{bmatrix} 
			0_3 & I_3
			\end{bmatrix} \mathbf{x} \left( t \right) \right] \cdot \left[ - K \mathbf{x} \left( t \right) \right]  \right\} \begin{bmatrix} 
			0_3 & I_3
			\end{bmatrix} \mathbf{x} \left( t \right) }{c^{2}}} - \frac{m_{0}}{\left[ 1 - \frac{\left| \begin{bmatrix} 
			0_3 & I_3
			\end{bmatrix} \mathbf{x} \left( t \right) \right|^2 }{c^2} \right]^{\frac{1}{2}}} K \mathbf{x} \left( t \right)
	\end{flalign}
	%\noindent\rule{\textwidth}{0.4pt}
\end{figure*}
while the relativistic output feedback controller can be derived as \eqref{long6} (see the last page).
\begin{figure*}
	\begin{flalign} \label{long6}
	\mathbf{u} \left( t \right) 
	= \frac{m_{0}}{\left[ 1 - \frac{\left| \dot{ \mathbf{y}} \left( t \right) \right|^2 }{c^2} \right]^{\frac{3}{2}}}{\frac{\left\{ \dot{ \mathbf{y}} \left( t \right) \cdot \mathbf{l} \left[  \mathbf{y} \left( t \right) \right]  \right\} \dot{ \mathbf{y}} \left( t \right) }{c^{2}}} - \frac{m_{0}}{\left[ 1 - \frac{\left| \dot{ \mathbf{y}} \left( t \right) \right|^2 }{c^2} \right]^{\frac{1}{2}}} \mathbf{l} \left[  \mathbf{y} \left( t \right) \right]
	\end{flalign}
	%\noindent\rule{\textwidth}{0.4pt}
\end{figure*}

%
%
%In addition, for output feedback, if $\mathbf{u} \left( t \right)$ is splitted into a component $\mathbf{u}_{\parallel} \left( t \right)$ parallel to $\mathbf{v} \left( t \right)$ and the other $\mathbf{u}_{\perp} \left( t \right)$ orthogonal to $\mathbf{v} \left( t \right)$ while $\mathbf{l} \left[  \mathbf{y} \left( t \right) \right]$ is also splitted into a component $\mathbf{l}_{\parallel} \left[  \mathbf{y} \left( t \right) \right]$ parallel to $\mathbf{v} \left( t \right)$ and the other $\mathbf{l}_{\perp} \left[  \mathbf{y} \left( t \right) \right]$ orthogonal to $\mathbf{v} \left( t \right)$, then it can be verified \cite{rindler2006relativity} that
%\begin{flalign} 
%\mathbf{u}_{\parallel} \left( t \right) 
%= \frac{m_{0}}{\left[ 1 - \frac{\left| \mathbf{v} \left( t \right) \right|^2}{c^2} \right]^{\frac{3}{2}}} \mathbf{l}_{\parallel} \left[  \mathbf{y} \left( t \right) \right],
%\end{flalign}
%while
%\begin{flalign} 
%\mathbf{F}_{\perp} \left( t \right) 
%= \frac{m_{0}}{\left[ 1 - \frac{\left| \mathbf{v} \left( t \right) \right|^2}{c^2} \right]^{\frac{1}{2}}} \mathbf{a}_{\perp} \left( t \right).
%\end{flalign}
%This means that it is as if the system manifests different inertial resistances (different ``masses"; see discussions in \cite{rindler2006relativity}) to the same force in different directions, depending on whether it is subjected to that force longitudinally or transversely.

\section{Conclusions}
% relativistic state feedback and relativistic output feedback, as well as relativistic PID control
In this note, we have introduced the state-space representation of the relativistic dynamics and investigated how to design the relativistic feedback controller based on the feedback linearization of the relativistic dynamics. 
%We have also examined concepts and methods such as relativistic controllability. 
Potential future research directions include analysis and design of state estimation (e.g., relativistic observer, relativistic filtering such as relativistic Kalman filter) of the relativistic dynamics.

\bibliographystyle{IEEEtran}
\bibliography{references}

\end{document}